\documentclass[letter,pre,showpacs,preprintnumbers,floatfix,twocolumn]{revtex4}

\usepackage{graphicx}
\usepackage{subfigure}
\usepackage[]{amssymb}
\usepackage[]{amsmath}

\begin{document}

\title{About the lifetime of a bouncing droplet}

\author{D. Terwagne, N. Vandewalle, and S. Dorbolo}

\affiliation{GRASP-Photop\^ole, Physics Department B5, University of Li\`ege, B-4000 Li\`ege, Belgium.}

\begin{abstract}
When a droplet is gently laid onto the surface of the same liquid, it stays at rest for a moment before coalescence.  The coalescence can be delayed and sometimes inhibited by injecting fresh air under the droplet. This can happen when the surface of the bath oscillates vertically, in this case the droplet basically bounces on the interface. The lifetime of the droplet has been studied with respect to the amplitude and the frequency of the excitation. The lifetime decreases when the acceleration increases.  The thickness of the air film between the droplet and the bath has been investigated using interference fringes obtained when the system is illuminated by low pressure sodium lamps.  Moreover, both the shape evolution and the motion of the droplet center of mass have been recorded in order to evidence the phase offset between the deformation and the trajectory. A short lifetime is correlated to a small air film thickness and to a large phase offset between the maximum of deformation and the minimum of the vertical position of the centre of mass.

\end{abstract}

\maketitle

\section{Introduction}

\begin{figure}[b]
\includegraphics[width=8cm]{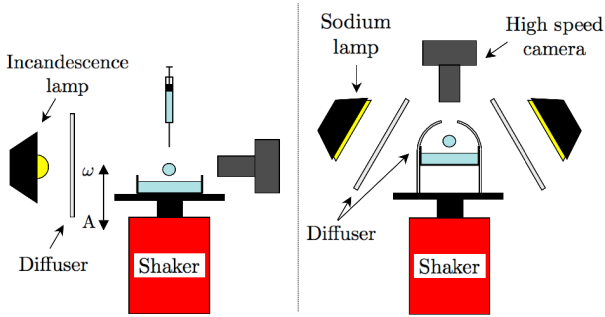}
\caption{\label{expset}(left) Experimental setup for the droplet trajectories and deformations characterizing composed by the electromagnetic shaker GW-V55, a plate on which the container filled with the oil bath is attached, an incandescence lamp and a high speed video camera. (rigth) In order to observe interference fringes, the container is lit homogeneously by three sodium low vapor pressure lamps. The droplet is filmed from the top using a high speed video camera.}
\end{figure}
The manipulation of tiny amounts of liquid plays a key role in any micro-device used in microfluidic applications. A droplet is often laid onto a substrate where it interacts with its environment. Some liquid may be lost onto the substrate or the separation with the surrounding liquid may be difficult. The idea is to manipulate the droplet without touching it \cite{pof} or in other words, to transport the droplet using another fluid, for example air.  Couder {\it et al} discovered that it is possible to make an oil droplet bouncing onto a vibrating bath of oil. Some air is squeezed between the droplet and the bath avoiding coalescence. The question about the stability of the bouncing droplet is naturally raised.  The aim of the present work is to discuss the lifetime of the bouncing droplet by studying its trajectory and the dynamics of the air film located between it and the bath.

When a droplet is gently laid onto the surface of the same liquid, it stays at the surface during the time that the air film located just below the droplet is drained out. As soon as the air film thickness reaches a critical value (about $100\,\rm{nm}$), the Van der Waals forces become of importance and the air film eventually collapses. Some non-coalescing drop systems have also been studied. It has been shown that, when a liquid-vapour CO$_2$ interface is shaken horizontally, droplets may be produced and set in motion on the interface \cite{Gonzales}. Sreenivas \textit{et al.} trapped droplets on an hydraulic jump \cite{Sreenivas}. Couder \textit{et al.} \cite{couder, couder2, couder3} have discovered that when the interface is vertically vibrated using an electromagnetic shaker (cf. Fig. \ref{expset}), the droplet may avoid the coalescence. Indeed, when the droplet bounces, the air film is renewed. The relevant parameter is the reduced acceleration $\Gamma$ defined as the ratio between the maximum acceleration $A\omega^2$ and the gravity constant acceleration $g$, $A$ and $\omega$ beeing respectively the amplitude and the pulsation of the oscillation. The bouncing can be obtained when $\Gamma$ is larger than a critical acceleration $\Gamma_{C}$ that scales as the square of the frequency $f=\omega/2\pi$ \cite{couder}. This criterion allows to determine whether the droplet bounces or not according to the couple of parameters $A$ and $\omega$.

\begin{figure}[b]
\includegraphics[width=8cm]{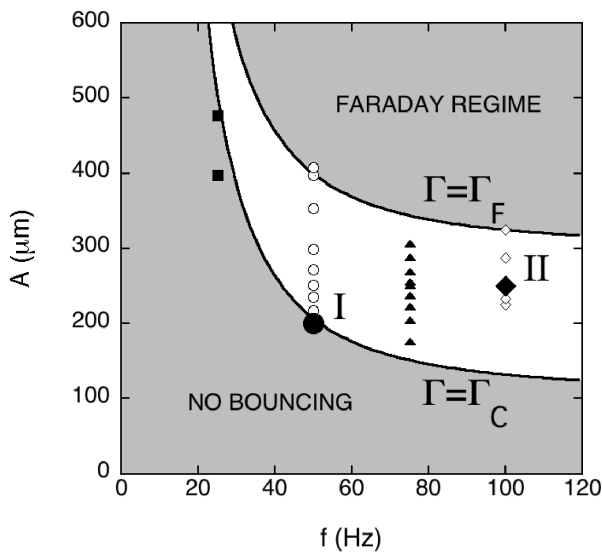}
\caption{\label{diagAw} Phase diagram amplitude-frequency for a droplet of $1.5\,\rm{mm}$ of diameter and made up of by oil of $50\,\rm{cSt}$. Top and bottom curves correspond to the Faraday $\Gamma_F$ and Couder $\Gamma_C$ acceleration thresholds respectively (see text). The data points represent the experimental conditions that have been investigated in the present work in order to cover at best the region where the droplet can bounce. Two large data points correspond to the cases I and II described in the text.}
\end{figure}

That phenomenon is perturbated at high accelerations when Faraday waves appear. This high acceleration limit $\Gamma_{F}$ depends on the viscosity of the oil in the bath. Considering a given droplet size and a given viscosity, the bouncing phenomenon may only occur in a delimited region of the ($A$, $f$) diagram, i.e. in between $\Gamma_{C}$ and $\Gamma_{F}$ (see Fig. \ref{diagAw}). 

A question remains about the stability of the bouncing droplet. Indeed, at a given frequency and for an acceleration larger than $\Gamma_C(\omega)$, the droplet may bounce for a time longer than a few days \cite{couder}. On the other hand, for a large frequency ($100\,\rm{Hz}$) and for $\Gamma > \Gamma_C(\omega)$, the droplet bounces only for a few seconds before coalescing. The aim of this work is to measure the lifetime of the bouncing droplets and to determine where, in the phase diagram $(A,\omega)$, they bounce the longest time.

Due to the large number of parameters : $A$, $f$, the radius of the droplet $R$, the viscosity of the oil $\eta_{o}$, some have been arbitrary fixed. The viscosity and the droplet size are fixed to $\eta_0=50\,\rm{cSt}$ and $2R=1.5\,\rm{mm}$ diameter respectively. In these conditions, the area we explored in the $A$-$f$ diagram is represented in Fig. \ref{diagAw} where the investigated set of parameters are represented. For each point, the lifetime has been measured. Then the air film dynamics has been studied by an interference fringe analysis.  Moreover, an optical tracking method has been used to determine the position and the deformation of the droplet with respect to the phase of the plate. 

After the description of the experimental setup (Sect. II), the lifetime analysis will be developped (Sect. III). Explanation will be found by studying the interference fringes (Sect. IV) as well as the trajectory and deformation of the droplet (Sect. V).

\section{Experimental setup}
A container is filled with silicon oil (Dow Corning 200 Fluid) having a viscosity of $50\,\rm{cSt}$. The plate is shaken according to a vertical oscillation of amplitude $A$ and of pulsation $\omega$ (Fig. \ref{expset} left). For each fixed frequency, the acceleration is tuned between $\Gamma_C$ and $\Gamma_F$ by changing the amplitude of the vibration. The droplets are produced by using a syringe and are gently dropped onto the surface of the bath. The lifetime is measured using a digital timer.

\begin{figure}[htbp]
\includegraphics[width=8cm]{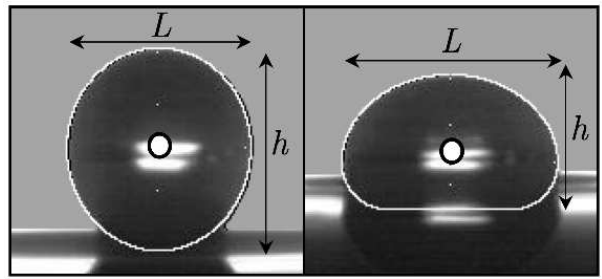}
\caption{\label{Ar} Snapshots of a droplet : (left) entirely visible, (right) at its maximum of deformation. The experimental contour of the droplet is represented by a white curve. The height $h$ and the width $L$ are indicated by the double arrows while the centre of mass is shown by the white circle.}
\end{figure}

The position and the shape of the droplets have been studied using image analysis. The motion of the droplet is recorded from the side using a fast video camera ($1000\,\rm{fps}$). Typical snapshots of a droplet are shown in Fig. \ref{Ar}. The one on the right presents a droplet squeezed by the plate. During a bounce, the deformation of the lower part and the upper part of the droplet is asymmetric. The evolution of its upper half part is easily determined by contour detection on images while the lower part of the droplet motion is not accessible part of the time. This difficulty occurs particularly when the droplet deforms the bath surface. However, as the total volume and the evolution of the upper part of the droplet are known, the evolution of its lower part may be deduced as soon as a hypothetical shape has been speculated. The lower part shape is modeled by a flat bottom surrounded by a quarter of a torus \cite{biance}. The shape is fitted to the experimental contour taking into account that the total volume must be conserved.  An example of this shape (white countour) is illustrated on Fig. \ref{Ar}. On the other hand, when the droplet is entirely visible, its shape is oblong, see Fig. \ref{Ar} (left). The lower part of the drop is supposed to behave like a half ellipsoid.  Eventually, the trajectory of the centre of mass is deduced as well as the evolution of the aspect ratio $A_r = h/L$, where $h$ and $L$ are respectively the height and the width of the droplet (Fig. \ref{Ar}).

In order to investigate the air film dynamics, three low pressure sodium lamps have been placed around the droplet (Fig. \ref{expset} left). That monochromatic light allows to visualize the interference fringes produced by the thin air film located between the droplet and the bath. A high speed video camera placed vertically above the plate has been used to record the motion of these fringes. The number of fringes gives information about the thickness of the air film \cite{klaseboer, zdravkov}.

\section{Lifetime of droplets}
\begin{figure}[htbp]

\includegraphics[width=8.5cm]{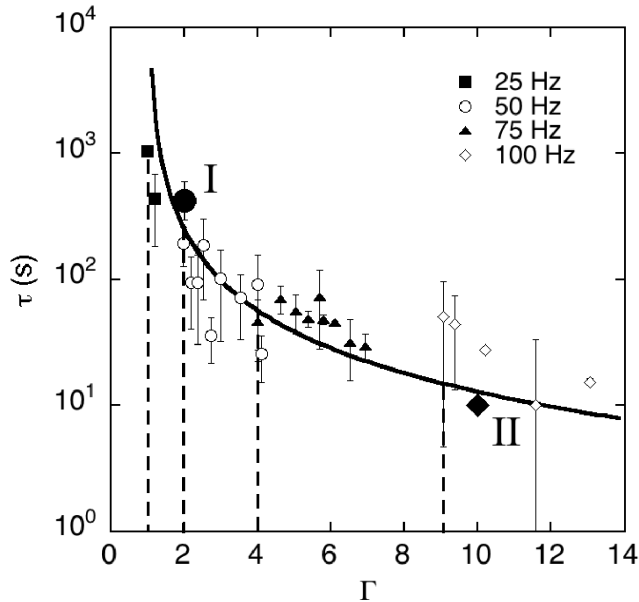}

\caption{\label{TV} Avereged lifetimes $\tau$ as a function of the reduced acceleration $\Gamma$. The symbols correspond to different considered frequencies (see legend). The vertical dashed lines represent the Couder thresholds according to the different frequencies. The continuous line is the fit from Eq. \ref{tauajust}.  The large points correspond to the case I and II described in the text.}
\end{figure}

Although Couder's criterion ($\Gamma > \Gamma_C$) is respected, we observed that bouncing droplets do not bounce for ever. Droplets lifetimes $\tau$ give us information on their stability beyond that threshold. Avereged lifetimes are represented in Fig. \ref{TV} with respect to the reduced acceleration $\Gamma$. The different symbols correspond to the considered frequency, i.e. the squares, circles, triangles and rhombuses are for 25, 50, 75 and $100\,\rm{Hz}$ respectively. The vertical dashed lines represent the Couder thresholds according to the four different considered frequencies.

The lifetime is found to drastically decrease with the reduced acceleration. One should note that, at 25 Hz, the lifetime is far below its real value. Indeed, the experiments were stopped when $\tau >  1000\,\rm{s}$. The droplet is qualified of stable beyond this arbitrary limit. A power law has been fitted (continuous line in Fig. \ref{TV}), namely
\begin{equation}
\tau \propto \frac{1}{(\Gamma-1)^{1.4}} \label{tauajust}
\end{equation}
A divergence is expected at $\Gamma = 1$, suggesting stability at very low frequencies. 

As far as the results corresponding to a given frequency are concerned, the maximum lifetime is found at the Couder threshold. Finally, the lifetimes corresponding to a same set of parameters are distributed along a broad distribution. This suggests that the mechanism explaining the life's duration of the droplet is probabilistic. 

In the next sections, two factors will be considered in order to explain the instability of the bouncing droplets. 

\section{Interference fringes}
\begin{figure}[h]
\vskip +5mm
\includegraphics[width=7.5cm]{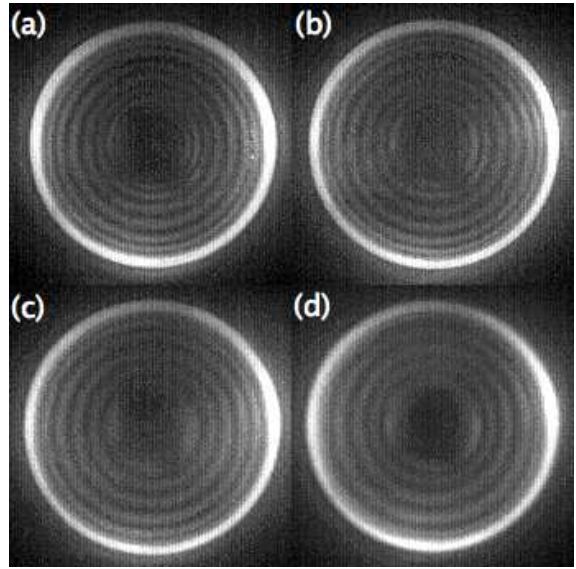}
\caption{\label{fringes}Interference fringes obtained when the air film located between the droplet and the bath is the most squeezed; the number of fringes is maximum. The different pictures have been obtained for the four frequencies (a) $26\,\rm{Hz}$, (b) $46\,\rm{Hz}$, (c) $74\,\rm{Hz}$, and (d) $98\,\rm{Hz}$.}
\end{figure}

One would expect that the instability of a bouncing droplet is due to the speed of the air drainage located between the droplet and the bath surface. In order to test this hypothesis, the bouncing droplet has been lit by sodium low pressure lamps. As this light may be considered monochromatic, interference fringes can be observed through the air gap separating the droplet and the bath. 

The interference fringes produced by the air film have been recorded with a high speed video camera. During a certain period, the direct observation of the fringes shows that the circular fringes are visible when the droplet deforms the surface of the bath. The fringes get closer and closer before separating and disappearing when the droplet takes off.  When a movie taken at the beginning of the bouncing is compared to another taken at the end of the bouncing life, no evolution is observed. The motion of the fringes remains periodical and identical during the whole life of the bouncing droplet whatever the frequency ! That observation means that the air film minimum thickness does not evolve. The death of the droplet cannot be explained by a progressive drainage of the air film. Taking this periodicity as an advantage, the observation frequencies have been chosen to obtain the maximum resolution time. Indeed, since the motion is periodical, an aliasing technique can be applied.

\begin{figure}[h]
\vskip +5mm
\includegraphics[width=8.5cm]{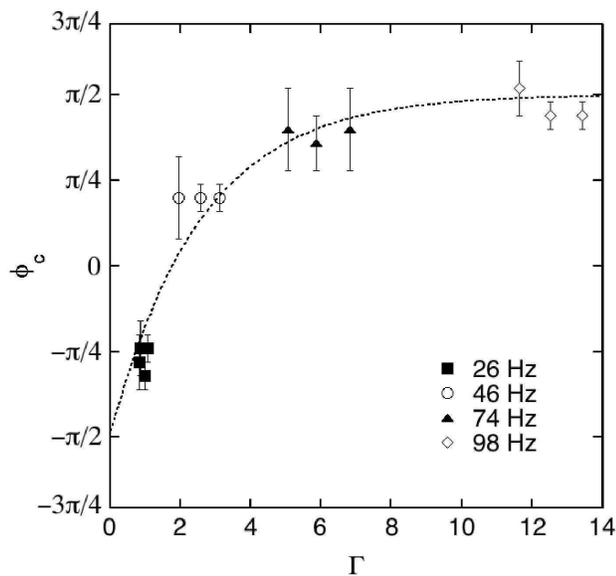}
\caption{\label{phic}Phases $\phi_c$ when the fringes are the closest as a function of the reduced acceleration $\Gamma$. The different frequencies considered have been represented by different symbols (see legend). The dashed curve is a guide for the eyes.}
\end{figure}

The movies allow to find the moment when the air film is the most squeezed, i.e. the phase $\phi_c$ at which the fringes are the closest. In Fig. \ref{fringes}, four pictures corresponding to this instant are compared for $f = 26\,\rm{Hz}$, (b) $46\,\rm{Hz}$, (c) $74\,\rm{Hz}$, and (d) $98\,\rm{Hz}$.  In Fig. 6, the phase $\phi_c$ is represented versus the acceleration $\Gamma$.  The phase is calculated considering that the plate position follows a sinusoidal law $A \sin (\omega t)$.  Two significant observations can be made. (i) The number of fringes decreases when the frequency increases. Their number is related to the thickness of the film \cite{couder}, a small number of fringes is characteristic of a thin film. That means that the minimum thickness of the air film is smaller at high frequency than at low frequency. (ii) The phase $\phi_c$ increases with the frequency and the acceleration (see Fig. \ref{phic}) and seems to saturate towards $\pi/2$, i.e. when the plate reaches its maximum position.  In the next section, an additional study will be performed between the maximum of squeezing and the trajectory of the bouncing droplet.

The probability of an air film to collapse is supposed to depend on its thickness. An activated process like an Arrhenius law may be imagined \cite{SD1,SD2}. Under this hypothesis, the lifetime is determined by the minimum thickness that the air film reaches during one period.  In order to explain why the minimum air film thickness changes with the frequency and the acceleration, the trajectory of a bouncing droplet will be analyzed in the next section.

\section{Trajectories and deformations}

In this section, the droplet deformations are finely studied for two very different cases : (i) case I : a bouncing droplet characterized by a lifetime of about 420 s ($A = 200\,\mu\rm{m}$ and $f = 50\,\rm{Hz}$) and (ii) case II : a bouncing droplet characterized by a lifetime of about 10 s ($A = 250\,\mu\rm{m}$ and $f = 100\,\rm{Hz}$).  Both cases are represented by large symbols in Fig. 2 and Fig. 4.

\begin{figure}[h]
\includegraphics[width=8.5cm]{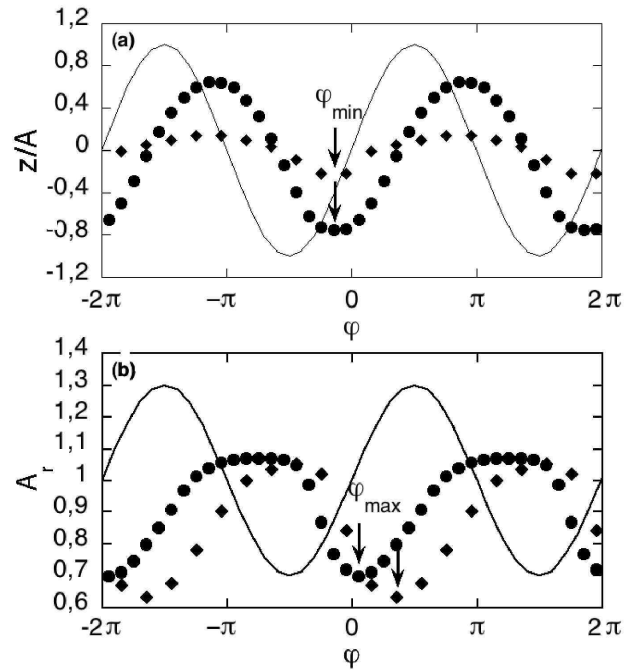}
\caption{\label{CMRa} Comparison of (a) the dimensionless position $z/A$ of the centre of mass and (b) the aspect ratio $A_r$ variation with respect to the reduced time ($\varphi = \omega t$) for cases I and II (see text). Cases I and II are represented by $\bullet$ and {\footnotesize $\blacklozenge$} respectively. The continuous curve gives an indication concerning the phase of the plate.}
\end{figure}

In Fig. \ref{CMRa}, cases I ($\bullet$) and II ({\footnotesize $\blacklozenge$}) are compared along two parameters. (i) Fig. \ref{CMRa}a compares the positions of the centre of mass while (ii) Fig. \ref{CMRa}b compares the variations of the aspect ratios $A_r$ with respect to the phase $\varphi$ defined by $\varphi = \omega t$ where $t$ is the time. Let us remember that a value $A_r=1$ corresponds to a spherical droplet. The vertical position $z$ of the centre of mass has been normalized by the amplitude $A$ of the excitation. The continuous curves represent the position of the plate in arbitrary units in order to visualize its phase. The normalization of the time is justified by the fact that the motion is periodic from the birth till the death of the bouncing droplet. As for the interference fringes, no deviation from these curves $z/A(\varphi)$ and $A_r(\varphi)$ is observed during their life. 

Two particular phases will be discussed : (i) the phase $\varphi_{min}$ at which the vertical position of centre of mass is minimum (see arrows Fig. \ref{CMRa}.a) and (ii) the phase $\varphi_{max}$ at which the deformation is maximum, i.e. $A_r$ is minimum (see arrows Fig. \ref{CMRa}.b). The normalized position of the droplet is characterized by a larger amplitude in the low frequency case than in the high frequency one. However, the phase $\varphi_{min}$ is nearly the same in both cases. These particular phases are indicated by arrows in Fig. \ref{CMRa}a. It is noticeable that the trajectory in case I is more symetrical than in case II.

The variation of the aspect ratio has nearly the same amplitude in both cases while the phases $\varphi_{max}$ are however very different. The bouncing droplet is deformed during a smaller period in case I than in case II. Indeed, the aspect ratio $A_r$ in case I is close and even above the unity during a longer period of time. This means that the time of interaction between the plate and the droplet is smaller than in case II. In that later case, the droplet is continuously deformed by the plate. 

In view of these results, a long lifetime is characterized by a large amplitude motion of the center of mass and a small phase difference between the maximum deformation and the minimum of the center of mass. This phase difference can be explained by considering the inertia and the viscosity of the droplet. When the forced deformation rate is too large (for high frequency), the droplet cannot follow the motion of the plate. A delay exists between the motion of the plate and the deformation that is slowed down by the inertia and the viscosity. It results in a stronger squeezing of the air film as observed in fringe figures. The ideal situation (for a long lifetime) is obtained when $\varphi_{min} \approx \varphi_{max}$.

\section{Conclusions}
We have studied the stability of bouncing droplets made up of silicon oil of $50\,\rm{cSt}$. The size of the droplet is $1.5\,\rm{mm}$. For this system, the lifetime of a bouncing droplet has been found to decrease with $\Gamma$ and to diverge for $\Gamma = 1$ and $f = 25\,\rm{Hz}$. Moreover, the distribution of the lifetimes for the same set of parameters suggests a probabilistic mechanism for the coalescence moment of the bouncing droplet. 

The motion of the air film and of the droplet has been tracked and recorded using image analysis. Both motions have been found to be periodical and identical during the whole life of the bouncing droplet. Such periodical motions indicate that the speed of the drainage cannot explain the decrease of the lifetime at high frequencies. On the other hand, the interference fringe analysis allows to determine that minimum thickness of the air film located between the droplet and the bath is smaller for high frequencies than for smaller frequencies.  The film is more fragile at high frequencies and has consequently a larger probability of collapsing each time that the air film is squeezed by the plate.

As for the measurement of the trajectory, the phase difference between the maximum of deformation and the minimum of the position of the center of mass has been found to be a key explaining the squeezing difference at low and high frequencies and therefore the difference of lifetime between the droplets. When this dephasage is close to zero and when the amplitude of the motion is large, the lifetime is increased. At high frequency, the phase difference increases and the interaction between the plate and the droplet is increased; the air film thickness is then decreased.

SD thanks the FNRS for its financial support. Exchanges between laboratories have been financially helped by the COST action P21. T. Gilet is acknowledged for fruitful discussions.  Part of this work has been supported by Dow Corning (J-P Lecomte, Seneffe). 


\begin{thebibliography}{99}
\bibitem{pof} N. Vandewalle, D. Terwagne, K. Mulleners, T. Gilet and S. Dorbolo, Phys. Fluids \textbf{18}, 091106Ê(2006) 
\bibitem{Gonzales} W. Gonz{\'a}lez-Vi{\~n}as and J. Sal{\'a}n, Europhys. Lett. \textbf{41}, 159 (1998)
\bibitem{Sreenivas} K. R. Sreenivas, P. K. De and J. H. Arakeri, J. Fluid Mech. \textbf{380}, 297 (1999) 
\bibitem{couder} Y. Couder, E. Fort, A. Boudaoud and C. H. Gautier, Phys. Rev. Lett. \textbf{94}, 177801 (2005) 
\bibitem{couder2} S. Proti\`ere, A. Boudaoud and Y. Couder, J. Fluid Mech. \textbf{554}, 85 (2006) 
\bibitem{couder3} Y. Couder and E. Fort, Phys. Rev. Lett. \textbf{97}, 154101 (2006) 
\bibitem{biance} A-L. Biance, C. Clanet and D. Qu\'er\'e, Phys. Fluids \textbf{15}, 1632 (2003) 
\bibitem{klaseboer} E. Klaseboer, J.Ph. Chevaillier, C. Gourdon and 0. Masbernat, J. Coloid and Interface Sci. \textbf{229}, 274 (2000)
\bibitem{zdravkov} A. N. Zdrakov, G. W. M. Peters and H. E. H. Meijer, J. Coloid and Interface Sci. \textbf{266}, 195 (2003)
\bibitem{SD1} S. Dorbolo, H. Caps and N. Vandewalle, New J. Phys. \textbf{5}, 161 (2003)
\bibitem{SD2} S. Dorbolo, E. Reyssat, N. Vandewalle and D. Qu\'er\'e, Europhys. Lett. \textbf{69}, 966 (2005)

\end{thebibliography}

\end{document}